\documentclass[]{relaxed_system_lab}

\usepackage[utf8]{inputenc}
\usepackage[T1]{fontenc}
\usepackage{geometry}
\usepackage{amsmath,amssymb}  %
\usepackage{charter}

\usepackage[toc,page,header]{appendix}

\usepackage{minitoc}
\usepackage{cleveref} 
\usepackage{subcaption}
\usepackage{booktabs}
\usepackage{graphicx}
\usepackage{pgfplots}
\usepackage{pgfplotstable}
\usepackage{xcolor}
\usepackage{CJKutf8}
\usetikzlibrary{patterns}
\usepackage{float}
\usepackage{caption}
\usepackage{bm}

\usepackage{soul} %
\usepackage{algorithm}      %
\usepackage{algpseudocode}  %
\usepackage{parskip}

\newcommand{\name}{\textsc{DafnyComp}\xspace}
\newcommand{\company}[1]{\textbf{#1}}
\newcommand{\model}[1]{\textsc{#1}}

\usepackage{hyperref}
\usepackage{url}
\usepackage{graphicx}
\usepackage{amsmath}
\usepackage{amssymb}
\usepackage{listings}
\usepackage[dvipsnames]{xcolor}
\usepackage{tikz}
\usepackage{subcaption}
\usepackage{enumitem}
\usepackage{parskip}
\usepackage{siunitx}
\usepackage{longtable}
\usepackage{multirow}
\usepackage{xspace}
\usepackage{tcolorbox}

\definecolor{codegreen}{rgb}{0,0.6,0}
\definecolor{codegray}{rgb}{0.5,0.5,0.5}
\definecolor{codepurple}{rgb}{0.58,0,0.82}
\definecolor{backcolour}{rgb}{0.95,0.95,0.92}
\definecolor{amber}{rgb}{1.0, 0.75, 0.0}
\definecolor{slategray}{RGB}{112,128,144}
\definecolor{hldafny}{rgb}{1,1,0.5}
\definecolor{teal}{rgb}{0.0, 0.5, 0.5}
\definecolor{amethyst}{rgb}{0.6, 0.4, 0.8}

\title{Local Success Does Not Compose:\\
Benchmarking Large Language Models for Compositional Formal Verification}

\author{Xu Xu$^{1,*}$}
\author{Xin Li$^{2,*}$}
\author{Xingwei Qu$^3$}
\author{Jie Fu$^4$}
\author{Binhang Yuan$^{1,\dagger}$}

\affiliation{$^1$HKUST}
\affiliation{$^2$NTU}
\affiliation{$^3$UoM}
\affiliation{$^4$Shanghai AI Lab}

\abstract{
Despite rapid advances in code generation, current Large Language Models (LLMs) still lack an essential capability for reliable and verifiable code generation: \emph{compositional reasoning} across multi-function programs. To explore this potential and important gap, we introduce \name, a benchmark designed to systematically evaluate LLMs on the generation of compositional specifications in Dafny. Unlike prior benchmarks that primarily target single-function annotation, \name focuses on programs composed of multiple interacting functions with necessary data dependencies, requiring LLMs to produce specifications that ensure correctness across component boundaries. Our benchmark comprises 300 automatically synthesized programs, each carefully constructed by combining 2--5 originally independent functions in a chain-based manner through LLM-driven synthesis. We evaluate LLMs from five leading research groups that represent the current frontier of reasoning-centric AI, including the \model{GPT}, \model{Claude}, \model{Gemini}, \model{DeepSeek}, and \model{Qwen} families. Our results reveal a striking dichotomy: while LLMs achieve both high syntax correctness ($>$99\%) and moderate verification rates ($>$58\%) in prior single-function benchmarks, they exhibit degraded syntax correctness (95.67\%) and a catastrophic verification failure (3.69\%) in \name's compositional tasks---a 92\% performance gap. Even the most powerful LLM achieves only 7\% verification at Pass@8, with most LLMs below 2\%. Further analysis reveals that LLMs systematically fail at cross-functional reasoning through three primary failure modes: \textit{specification fragility} (39.2\%), \textit{implementation-proof} misalignment (21.7\%), and \textit{reasoning instability} (14.1\%). These failures clearly reveal the absence of compositional reasoning capabilities in current LLMs. \name thus establishes a diagnostic benchmark for tracking progress in verifiable code generation with LLMs, highlighting that the path from local to compositional verification remains largely uncharted. 

\begin{tabular}{@{}l l@{}}  
    \textbf{Project homepage:} \url{https://DafnyComp.github.io}
\end{tabular}
}

\begin{document}

\maketitle

\let\thefootnote\relax\footnotetext{\textsuperscript{*}Equal contribution. \textsuperscript{\dag}Correspondence to: Binhang Yuan \texttt{<biyuan@ust.hk>}.}

\section{Introduction}
Large language models (LLMs) have transformed software development through their remarkable code generation capabilities, enabling developers to produce complex programs from natural language descriptions~\citep{chen2021evaluating, austin2021program}. These advances have driven widespread adoption of programming assistants and development environments, fundamentally transforming how modern software is developed. As LLM-generated code becomes increasingly integrated into production systems, a critical question emerges: \textit{how to ensure the correctness of automatically synthesized programs}. Unlike human-written code that can be manually reviewed and tested, the scale and complexity of LLM outputs demand systematic approaches to verification that go beyond traditional debugging methods. On the other hand, conventional testing provides only partial confidence and cannot rule out rare corner cases or subtle specification mismatches.

\begin{figure}[t]
    \centering
    \includegraphics[width=\textwidth]{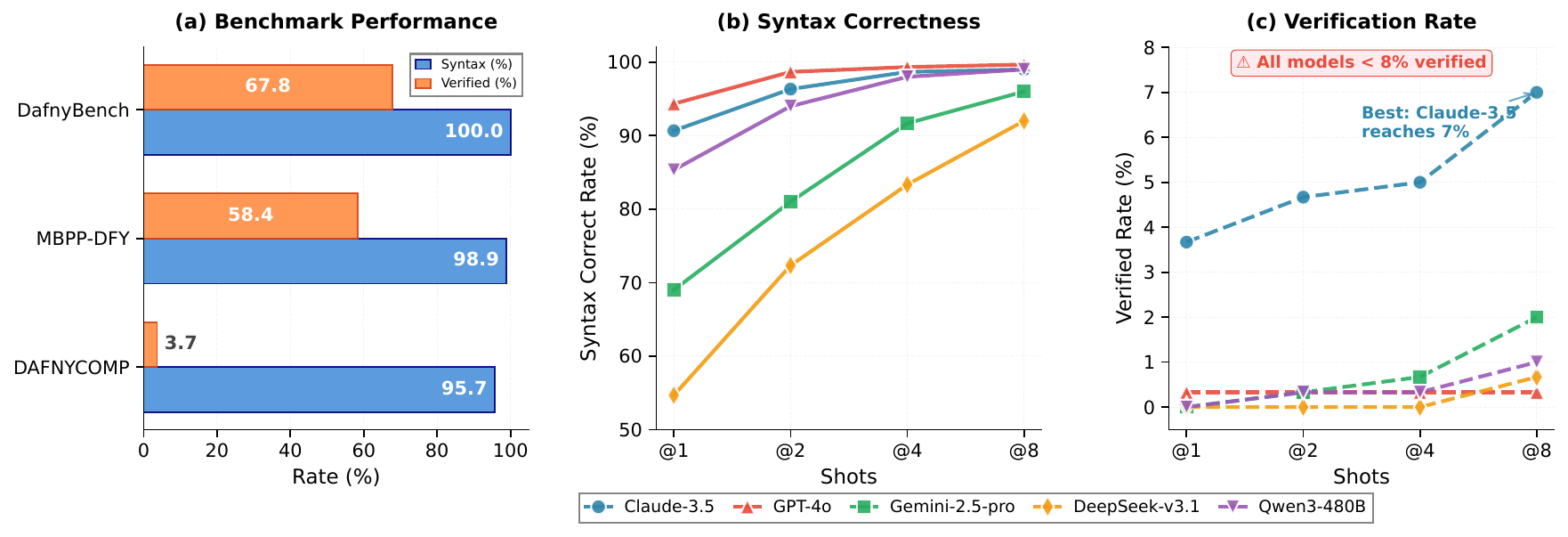}
    \caption{\textbf{The formal verification gap: high syntax success versus low verification rates.} 
    (a)~Benchmark performance reveals a dramatic gap between syntax correctness and verification success, 
    with \name showing a 92\% drop from 95.67\% to 3.69\%. 
    (b)~All models converge to high syntax correctness at @8 shots, with performance ranging from 92\% to 99\%. 
    (c)~Verification rates remain critically low ($<$8\%) across all models despite increased sampling, with Claude-3.5 achieving the highest rate at only 7\%.}
    \label{fig:overview}
    \vspace{-1.0em}
\end{figure}

Formal verification provides a principled solution to this challenge by offering mathematical guarantees of program correctness through rigorous specification and proof techniques. Programming languages like Dafny enable developers to express precise contracts---preconditions, postconditions, and invariants---that can be mechanically verified against implementations~\citep{leino2010dafny_tour}. However, the adoption of formal verification has historically been constrained by what we often refer to as the ``specification bottleneck'': writing comprehensive annotations not only demands specialized expertise but also produces specification code that is comparable in size to the implementation itself~\citep{leino2017compositional, loughridge2024dafnybench_synth}. Recent research has explored the use of LLMs to automate this specification generation process, demonstrating promising results where models can complete missing annotations for individual functions and achieve moderate verification success rates~\citep{loughridge2024dafnybench, yan2025reform}. However, current work in this area often suffers from a critical limitation: existing benchmarks, such as \textsc{DafnyBench}, primarily evaluate annotation completion within isolated functions~\citep{loughridge2024dafnybench}, failing to address the compositional reasoning ability required for real-world, sophisticated software systems, where correctness emerges from complex interactions between multiple components~\citep{keysers2020compositional}.

To fill this gap, we introduce \name, the first benchmark explicitly designed to evaluate the generation of compositional specifications for programming languages equipped with formal verification. Concretely, we make the following contributions:

\textbf{\underline{Contribution 1.}} To address the limitations of prior verification benchmarks, we present \name, a new benchmark explicitly designed for compositional formal verification (\S\ref{sec:benchmark_constrcut}). Unlike existing datasets such as \textsc{DafnyBench}~\cite{loughridge2024dafnybench} that focus on specification generation for isolated single functions, \name evaluates LLMs on programs composed of multiple interacting functions with real data dependencies. The benchmark consists of 300 Dafny programs synthesized by combining 2–5 independent functions, forcing models to reason across function boundaries to ensure end-to-end correctness. Note that our design is the first to require actual compositional reasoning in specification generation, bridging a critical gap left by prior benchmarks and reflecting the complexities of real-world software systems.

\textbf{\underline{Contribution 2.}}  We comprehensively evaluate 13 state-of-the-art LLMs on the constructed \name (\S\ref{sec:exp}) benchmark, including advanced models like \textsc{GPT-4o}, \textsc{Claude 3.5}, \textsc{Gemini 2.5}, \textsc{DeepSeek-v3.1}, and \textsc{Qwen3-coder}. The results reveal a dramatic collapse in verification performance despite high syntactic accuracy: while the models produce syntactically correct code for approximately 95.7\% of the tasks, only 3.7\% of their outputs actually pass the formal verifier. This staggering ~92\% gap between syntax success and semantic correctness persists across all model families, prompt settings, and sampling strategies. Even with up to 8 attempts per problem, the best model attains only around 7\% verification success, indicating that increasing sampling or context does not remedy the fundamental limitation.

\textbf{\underline{Contribution 3.}} We carefully analyze the failure cases in the benchmark, which pinpoints three primary failure modes underlying this breakdown (\S\ref{sec:insight}), highlighting systemic obstacles to compositional reasoning in current LLMs: 

\begin{itemize}[leftmargin=*]
\item (\underline{\textbf{i}}) \textit{Specification fragility}: we observe the brittleness of generated specifications wherein small omissions, over-/under-strengthening, or inconsistent framing (\texttt{reads}/\texttt{modifies}) clauses can invalidate downstream proofs. Concretely, in compositional settings, a missing or slightly weaker postcondition at one stage can fail to imply a callee’s precondition, triggering a domino effect along the call chain even when each component appears locally reasonable.

\item (\underline{\textbf{ii}}) \textit{Implementation–proof misalignment}: We identify inconsistencies between the produced code and its associated specifications or proofs, indicating that models often generate implementations and annotations via largely independent pathways. Typical symptoms include plausible-but-false loop invariants, contradictory \texttt{requires}/\texttt{ensures} obligations, or termination metrics that do not match control flow, any of which cause verification to fail despite syntactic well-formedness.

\item (\underline{\textbf{iii}}) \textit{Reasoning instability}: we witness a tendency to lose the inductive thread of the argument over multiple steps, leading to invariants that are not preserved, incomplete coverage of cases, or missing well-founded \texttt{decreases} measures. These errors are most evident in composition, where maintaining complex state relationships across iterations and function boundaries is crucial for end-to-end correctness. 
\end{itemize}

These failure modes were pervasive in the models’ outputs, revealing a fundamental absence of robust compositional reasoning capabilities. By identifying these issues, \name serves as a diagnostic benchmark for the community, enabling systematic measurement of progress toward LLMs that can verify complex multi-component programs.

\section{Related Work}

\textbf{Formal Verification Benchmarks.}
Existing benchmarks for verifiable code generation can be categorized into two types. Single-function benchmarks, such as \textsc{DafnyBench}~\citep{loughridge2024dafnybench} and \textsc{MBPP-DFY}~\citep{misu2024towards}, evaluate annotation completion within isolated methods, achieving moderate success rates (50-60\%) but failing to capture inter-function dependencies. Interactive theorem proving benchmarks (miniCodeProps~\citep{lohn2024minicodeprops}, FVAPPS~\citep{dougherty2025proving}) target proof synthesis in systems like Lean~\citep{de2015lean} but require extensive manual validation and remain disconnected from practical programming. \name bridges this gap by evaluating compositional specification generation---a prerequisite for scaling verification beyond toy programs to production systems. Unlike prior work, we explicitly construct multi-function programs with data dependencies, exposing the compositional reasoning deficit in current models (see Appendix~\ref{appendix:comparison} for detailed comparison).

\textbf{Dynamic Benchmark Generation.}
Static benchmarks suffer from contamination and overfitting~\citep{hu2025dynacode,zhang2024darg}. Dynamic generation techniques---creating new tasks, transforming problems, or perturbing reasoning structures~\citep{zhu2024dyval,zhu2023dyval}---show promise in mathematics and logic but neglect formal verification's unique demands: specifications must be syntactically valid, semantically precise, and correct across all execution paths. Our synthesis pipeline addresses this by generating verifiable multi-function Dafny programs through controlled composition, ensuring both novelty and correctness while maintaining the semantic complexity that exposes compositional reasoning gaps.

\textbf{Compositional Reasoning in LLMs.}
The ability to systematically combine simpler units into correct larger structures remains a frontier challenge~\citep{li2024understanding,dziri2023faith}. While progress exists in natural language and symbolic domains, formal verification imposes stricter demands: specifications must preserve invariants across components and ensure global correctness. Existing training paradigms favor pattern matching over principled proof construction. By requiring models to generate specifications bridge function boundaries with explicit data dependencies, \name provides the first diagnostic benchmark for compositional reasoning in formal verification.

\section{Benchmark Construction}
\label{sec:benchmark_constrcut}

\begin{figure}[ht]
    \centering
    \includegraphics[width=\linewidth]{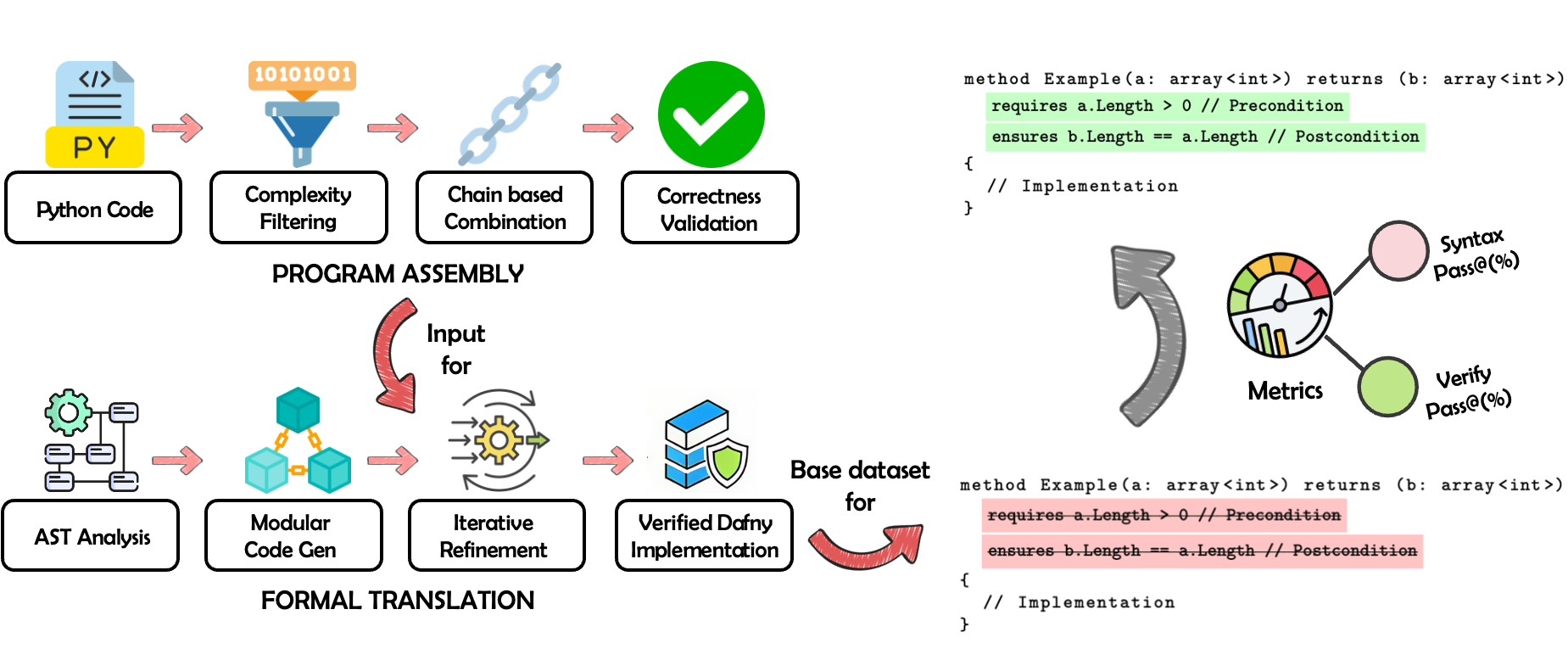}
    \caption{Two-stage benchmark synthesis: (1) Assembly combines independent Python functions with controlled data flow, ensuring algorithmic complexity while maintaining tractability; (2) Formal translation converts to verified Dafny through incremental AST-guided transformation.}
    \label{fig:pipe}
    \vspace{-1.0em}
\end{figure}

\name synthesizes 300 verified multi-function programs through a two-stage pipeline (Figure~\ref{fig:pipe}), including program assembly (\S\ref{sec:assembly}) and formal translation (\S\ref{section:trans}), which bridges the gap between practical Python implementations and verification-oriented Dafny specifications. 
Within this pipeline, program assembly ensures the construction of compositional Python programs with functional correctness, while specification translation with refinement ensures the quality and reliability of the resulting data. 
We also provide format details of evaluation tasks (\S\ref{sec:task_format}) and the key characteristics of the benchmark (\S\ref{sec:task_summary}).

\subsection{Program Assembly}
\label{sec:assembly}

We construct compositional programs by systematically combining functions from \textsc{LeetCodeDataset}~\citep{xia2025leetcodedataset}, selected for their algorithmic depth and verification challenges.

\textbf{Function Selection.} We filter the corpus using McCabe's cyclomatic complexity~\citep{mccabe1976complexity} as a proxy for verification difficulty, retaining only functions with complexity $>$5 (around top 30\% of the dataset) and at least 10 lines of code. This threshold ensures non-trivial control flow---loops with complex termination conditions, nested conditionals, recursive patterns---that stress specification generation. For tractability, we restrict to single-input/single-output functions, yielding 1,847 candidate functions.

\textbf{Compositional Strategy.} Following~\citet{hu2025dynacode}, we employ chain-based composition where each function's output feeds the next's input, creating explicit data dependencies. While more complex call graphs (trees, DAGs) are theoretically richer, empirical trials showed synthesis success drops from 47\% (chains) to $<$8\% (arbitrary graphs) while providing no additional diagnostic value---the chains suffice to expose compositional failures. After composition, we further identify the minimal set of shared import dependencies across the combined Python functions. This step is essential because the original \textsc{LeetCodeDataset} often relies on broad \texttt{import *} statements, which obscure library ownership and names. Without explicit mappings, Dafny cannot interpret external libraries, preventing the synthesis of intermediate functions to replace missing third-party features. We generate programs with 2–5 functions, exploring multiple permutations since function ordering affects both data flow and verification complexity.

\textbf{Validation Pipeline.} After composition, the resulting Python code is subjected to a three-stage validation pipeline, which filters candidates before their use in Section~\ref{section:trans}.

\begin{itemize}[leftmargin=*]
\item 
(\underline{\textbf{i}}) \textit{Type checking via constraint propagation}: We statically infer candidate types and shapes for each function’s inputs and outputs and propagate these constraints along the composition chain. This pass rejects compositions with incompatible interfaces (e.g., scalar-sequence or element-type mismatches) and flags violations of simple value constraints inferred from guards (such as non-negativity or length bounds). The result is a set of compositions whose interfaces are consistent end-to-end, providing a reliable basis for subsequent translation and verification.

\item (\underline{\textbf{ii}}) \textit{Formatting standardization}: We apply a deterministic rewriter, implemented with tools such as \texttt{Black} and \texttt{isort}, to normalize code style, including indentation, whitespace, line breaking, and import organization. Canonicalizing these incidental variations yields stable, diff-friendly artifacts and reduces prompt variance in later stages. This step preserves semantics while producing uniform program layouts that are easier to parse, translate, and verify.

\item (\underline{\textbf{iii}}) \textit{Test validation}: We execute each composed program against the reference unit tests from \textsc{LeetCodeDataset} to confirm functional correctness and basic executability. Programs that raise runtime exceptions, fail assertions, or produce incorrect outputs are discarded, ensuring only behaviorally sound compositions advance. This filtering isolates verification challenges to specification and reasoning rather than implementation errors during the Dafny translation stage.

\end{itemize}

Following this procedure, we obtain 1,200 valid Python programs with 2–5 functions, which will be used in the next stage.

\subsection{Formal Translation}
\label{section:trans}

We translate validated Python compositions into Dafny implementations with formal guarantees, focusing on the verification-oriented aspects of the benchmark here.

\textbf{Translation Challenges.} Direct end-to-end translation from Python to Dafny proved largely ineffective, with empirical success rates below 5\%. The core difficulty lies in Dafny’s demand for explicit specifications, invariants, and termination arguments—semantic elements absent in Python. This semantic gap makes single-pass translation infeasible for non-trivial programs.

\textbf{Incremental Pipeline.} Inspired by~\citet{wen2024enchanting}, we adopt an incremental approach: the abstract syntax tree (AST) of each Python program is decomposed into function- or control-structure–level fragments. Each fragment is translated into Dafny and immediately verified, localizing errors to the smallest possible unit. Verified fragments are then progressively reassembled according to the AST hierarchy, culminating in a complete Dafny program. Importantly, although translation proceeds incrementally, the Python program must be composed in its entirety before it can be executed. Whole-program composition in Python provides two benefits: (\underline{\textbf{i}}) Python’s explicit AST nodes and mature tooling make program assembly more reliable and transparent; and (\underline{\textbf{ii}}) having a coherent Python blueprint ensures that the incremental Dafny translation preserves global logical relationships, rather than producing isolated fragments that fail to compose. Thus, whole-program composition and incremental translation are complementary design choices. To further improve reliability, each candidate Dafny program undergoes up to ten refinement iterations, where specifications are strengthened in response to verifier feedback (e.g., adding loop invariants, refining postconditions, or inserting assertions). The entire synthesis and refinement process is carried out by \textsc{Claude‑4‑Sonnet‑20250514}, with the exact prompts provided in Appendix~\ref{appendix:prompt_syn}.

In total, the pipeline ultimately yields 564 verified Dafny programs from 1,200 attempts (corresponding to an overall 47\% success rate). Translation synthesis errors primarily arise from incomplete specifications (31\%), type inference errors (22\%), timeouts (18\%), and irreconcilable semantic gaps (29\%). From these, we retain 300 programs carefully balanced across complexity levels (100 each with 2–3, 3–4, and 4–5 functions). To ensure evaluation integrity, we conduct a thorough contamination analysis against \textsc{MBPP-DFY}~\citep{misu2024towards}, which is similarly synthesized from Python code as a Dafny benchmark dataset. The results in Appendix~\ref{sec:contamination} provide strong evidence that our test set is indeed free from bias due to data overlap.

\subsection{Evaluation Task Format}
\label{sec:task_format}

We adopt a specification reconstruction task inspired by~\citet{loughridge2024dafnybench}. Still, with a crucial difference: rather than removing all \texttt{assert} and \texttt{invariant} statements, we strip away the contract clauses (\texttt{requires}, \texttt{ensures}, \texttt{reads}, \texttt{modifies}, \texttt{decreases}) that appear before the opening brace of each \texttt{method} or \texttt{function}. LLMs to be evaluated are then required to regenerate these specifications to enable verification. This design isolates the challenge of reconstructing cross-function contracts from implementation concerns, focusing evaluation on whether models can generate specifications that capture emergent correctness properties across component boundaries. Unlike annotation completion tasks that permit purely local reasoning, our multi-function programs require understanding how data flows and invariants propagate through compositions. We employ a unified prompt across all evaluations (see Appendix~\ref{appendix:prompt}).

\subsection{Benchmark Statistical Summary}
\label{sec:task_summary}

The resulting benchmark comprises 300 mechanically verified Dafny test cases that jointly capture three key dimensions: compositional complexity from function-to-function call dependencies, algorithmic diversity across multiple categories, and verification challenges arising from the increased specification burden.

\textbf{Compositional Complexity.} Each program contains 2–5 functions (mean = 3.2) with an average of 8.4 cross-function data dependencies, requiring models to reason about specification alignment across component boundaries. Unlike single-function benchmarks where specification generation is largely local, our programs demand that preconditions of called functions be implied by postconditions of their callers—a requirement that introduces cascading verification challenges when specifications fail to propagate correctly.

\textbf{Algorithmic Diversity.} The benchmark spans 15 algorithmic categories with balanced representation: dynamic programming (18\%), string manipulation (20\%), number theory (15\%), and graph algorithms (12\%) constitute the primary categories, with the remainder distributed across sorting, searching, and combinatorial problems. Beyond the balance of individual categories, diverse permutations and combinations of these types yield composed programs with more intricate, layered structures. Consequently, the target of composing function calls is reinforced by the characteristics of the source dataset (\textsc{LeetCodeDataset}~\citep{xia2025leetcodedataset}), which in turn induces composition at the level of algorithmic logic. This design ensures models must develop general compositional reasoning rather than memorizing category-specific patterns.

\textbf{Verification Challenges.} Every program is mechanically verified by Dafny 4.10.0, thereby providing ground-truth correctness. The median program requires 7 loop invariants and 4 assertions for verification, with 23\% demanding intricate termination arguments via \texttt{decreases} clauses—a 3.5× increase in annotation density compared to \textsc{DafnyBench}'s average of 2 per program. This added specification burden reflects the extra complexity of compositional verification, creating a graduated challenge that pinpoints where current models shift from local reasoning to compositional failure.

\section{Experimental Setup And Results}
\label{sec:exp}

In this section, we enumerate the evaluation metrics (\S\ref{sec:metric}), LLM model selection for the benchmark (\S\ref{sec:model_sel}), and the corresponding evaluation results (\S\ref{sec:results}).

\subsection{Metrics}
\label{sec:metric}
We evaluate two complementary aspects of model performance:

\begin{itemize}[leftmargin=*]
    \item \textbf{Syntax Correctness}: measures whether generated specifications parse successfully in Dafny. This baseline metric captures models' grasp of the formal language syntax.
    \item \textbf{Verification Rate}: measures the fraction of syntactically correct programs that pass Dafny's verifier—the ultimate test of semantic understanding. This metric is computed only over syntactically valid outputs, as verification requires parseable code.
\end{itemize}

Following~\citet{chen2021evaluating}, we report Pass@$k$ for $k \in \{1,2,4,8\}$, measuring the overall probability of successfully solving a problem within $k$ attempts. Pass@1 provides a strict zero-shot baseline of immediate reasoning ability, whereas larger $k$ values further exploit additional test-time compute to improve success on compositional tasks~\citep{snell2024scaling}. In this setting, Pass@8 is particularly informative for clearly distinguishing model robustness and adaptability.

\subsection{Model Selection}
\label{sec:model_sel}

We evaluate 13 frontier models spanning five architectural families, chosen for their demonstrated strength in code generation and general reasoning:

\begin{itemize}[leftmargin=*]
\item \textbf{OpenAI}: \model{GPT-4o}~\citep{hurst2024gpt}, \model{GPT-4.1}~\citep{openai_gpt41}, \model{o4-mini}~\citep{openai_o4mini}
\item \textbf{Anthropic}: \model{Claude-3.5-Sonnet}~\citep{anthropic_claude35}, \model{Claude-4-Sonnet}~\citep{anthropic_claude4sonnet}
\item \textbf{Google}: \model{Gemini-2.5-Pro}, \model{Gemini-2.5-Flash}~\citep{google_gemini25}
\item \textbf{DeepSeek}: \model{DeepSeek-R1}~\citep{guo2025deepseek}, \model{DeepSeek-V3}~\citep{liu2024deepseek}, \model{DeepSeek-V3.1}~\citep{deepseekv3.1}
\item \textbf{Alibaba}: \model{Qwen2.5-Coder-32B}~\citep{hui2024qwen2}, \model{Qwen3-Coder-480B}~\citep{qwen3coder2025}, \model{QwQ-32B}~\citep{qwq32b2025}
\end{itemize}

\begin{table}[t]
{
\centering
\caption{Model performance reveals high syntax mastery but catastrophic verification failure. While syntax correctness reaches 99\% with sufficient sampling, verification rates remain below 7\% even for the best models, exposing the compositional reasoning gap.}
\label{tab:model_perf_grouped}
\resizebox{\textwidth}{!}{%
\begin{tabular}{lcccccccc}
\toprule
\multirow{2}{*}{\textbf{Model}}
  & \multicolumn{4}{c}{\textbf{Syntax Correct Rate (\%)}} 
  & \multicolumn{4}{c}{\textbf{Verified Rate (\%)}} \\
\cmidrule(lr){2-5} \cmidrule(lr){6-9}
  & @1 & @2 & @4 & @8 & @1 & @2 & @4 & @8 \\
\midrule
\multicolumn{9}{l}{\company{OpenAI Models}} \\
\quad \model{gpt-4o}      & 94.33 & 98.67 & 99.33 & 99.67 & 0.33 & 0.33 & 0.33 & 0.33 \\
\quad \model{o4-mini}     & 80.00 & 92.67 & 98.00 & 99.00 & 0.00 & 0.00 & 0.67 & 0.67 \\
\quad \model{gpt-4.1}     & 59.00 & 69.67 & 79.33 & 86.33 & 0.00 & 0.00 & 0.00 & 0.00 \\
\midrule
\multicolumn{9}{l}{\company{Anthropic Models}} \\
\quad \model{claude-3.5-sonnet}$^*$ & 90.67 & 96.33 & 98.67 & 99.00 & 3.67 & 4.67 & 5.00 & 7.00 \\
\quad \model{claude-4-sonnet}$^\dagger$   & 95.67 & 97.33 & 98.00 & 98.33 & 2.33 & 3.00 & 3.00 & 3.33 \\
\midrule
\multicolumn{9}{l}{\company{Google Models}} \\
\quad \model{gemini-2.5-flash} & 54.00 & 64.00 & 81.00 & 89.67 & 0.00 & 0.00 & 0.00 & 0.00 \\
\quad \model{gemini-2.5-pro}   & 69.00 & 81.00 & 91.67 & 96.00 & 0.00 & 0.33 & 0.67 & 2.00 \\
\midrule
\multicolumn{9}{l}{\company{DeepSeek Models}} \\
\quad \model{deepseek-r1}   & 85.67 & 95.33 & 98.33 & 99.00 & 0.33 & 0.33 & 0.33 & 0.33 \\
\quad \model{deepseek-v3}   & 77.33 & 88.67 & 95.33 & 97.33 & 0.00 & 0.00 & 0.33 & 0.33 \\
\quad \model{deepseek-v3.1} & 54.67 & 72.33 & 83.33 & 92.00 & 0.00 & 0.00 & 0.00 & 0.67 \\
\midrule
\multicolumn{9}{l}{\company{Alibaba Models}} \\
\quad \model{qwen3-coder-480b-a35b-instruct} & 85.33 & 94.00 & 98.00 & 99.00 & 0.00 & 0.33 & 0.33 & 1.00 \\
\quad \model{qwen2.5-coder-32b-instruct}     & 62.00 & 74.67 & 85.00 & 89.00 & 0.00 & 0.33 & 0.33 & 0.67 \\
\quad \model{qwq-32b}               & 46.67 & 61.33 & 78.00 & 91.00 & 0.00 & 0.00 & 0.00 & 0.00 \\
\bottomrule
\end{tabular}%
}
}
\vspace{-1.5pt}
{\footnotesize $^*$claude-3.5-sonnet-20241022, $^\dagger$claude-4-sonnet-20250514}
\vspace{-1.0em}
\end{table}

\subsection{Results and Discussion}
\label{sec:results}

Table~\ref{tab:model_perf_grouped} presents our main experimental findings. We observe a systematic verification collapse across all evaluated models, with four interesting observations:

\textbf{Observation 1.} \textit{Universal verification failure despite syntactic mastery.} 
The most striking result is the consistent 92-percentage-point gap between syntax correctness and verification success across all models. At Pass@8, models achieve $\mu = 95.67\%$ (SD = 4.21\%) syntax correctness but only $\mu = 3.69\%$ (SD = 2.14\%) verification. This gap persists independent of: (i) model scale (480B vs 32B parameters, $p > 0.05$), (ii) training specialization (code-specific vs general-purpose), (iii) architectural family (dense, MoE, constitutional), and (iv) increased sampling (Pass@1 to Pass@8). The universality of this failure suggests a fundamental architectural limitation rather than an optimization or data issue.

\textbf{Observation 2.} \textit{Non-linear scaling reveals compositional breakdown.}
Comparing performance degradation from single-function (\textsc{DafnyBench}) to multi-function (\name) tasks reveals super-linear complexity scaling. With 3.2$\times$ increase in functions, we observe a 14.4$\times$ decrease in verification success (from $\sim$53\% to 3.69\%). This disproportionate degradation cannot be explained solely by additive difficulty. Instead, it suggests that specification requirements grow combinatorially with function composition—each function boundary introduces $O(n^2)$ potential specification dependencies that models fail to capture.

\textbf{Observation 3.} \textit{Sampling saturation indicates capability ceiling, not search limitations.}
The verification-sampling curve plateaus by Pass@4 for all models, with the marginal improvement from Pass@4 to Pass@8 averaging only 0.8\%. In contrast, syntax correctness continues improving (+7.3\% on average), demonstrating that models can explore the output space but cannot discover valid specifications. This divergent behavior between syntax and semantics strongly suggests that current architectures lack the inductive biases necessary for compositional reasoning, rather than merely requiring better search strategies or more compute.

\textbf{Observation 4.} \textit{Reasoning-specialized models show no clear advantage, thereby confirming architectural barriers.}
Models explicitly optimized for reasoning (\textsc{QwQ-32B} with chain-of-thought focus, \textsc{DeepSeek-R1} with reinforcement learning) perform no better than general-purpose models, with \textsc{QwQ-32B} still achieving 0\% verification even at Pass@8. The tight clustering of verification rates (coefficient of variation = 0.58) across diverse training objectives strongly indicates that compositional verification requires fundamentally different architectural primitives—not refinements of existing transformer-based reasoning. This null result is particularly informative: it clearly demonstrates that neither extended reasoning traces nor reward-based optimization can overcome the absence of compositional inductive biases.

\section{Failure Case Analysis and Discussion}
\label{sec:insight}

\begin{table}[ht]
\centering
\caption{Distribution of verification failure modes across 900 analyzed cases from \name. Categories determined through automated error analysis and manual validation on 10\% sample.}
\label{tab:error-summary}
\resizebox{\textwidth}{!}{
\begin{tabular}{lccc}
\toprule
\textbf{Failure Mode} & \textbf{Frequency} & \textbf{\% of Total} & \textbf{Primary Mechanism} \\
\midrule
Specification Fragility            & 353/900 & 39.2 & Contract propagation failure \\
Implementation--Proof Misalignment & 195/900 & 21.7 & Independent generation pathways \\
Reasoning Instability              & 127/900 & 14.1 & Inductive chain breakdown \\
Other (syntax, timeout, misc.)    & 225/900 & 25.0 & Various \\
\bottomrule
\end{tabular}
}
\end{table}

The significant gap between syntax correctness and ultimate verification success demands a clear mechanistic explanation. Through a systematic analysis of 900 observed verification failures across three representative model families, we identify several distinct failure modes that reveal fundamental limitations in how transformers process compositional specifications. Table~\ref{tab:error-summary} presents the overall distribution of these failures, which we analyze in detail below.

\subsection{Specification Fragility: The Domino Effect}
\textit{Specification fragility}, the inability to generate contracts that remain valid across function compositions, constitutes the plurality of failures. Consider a representative case from our benchmark: a \texttt{digitSum} function correctly implemented but missing the postcondition \texttt{ensures result >= 0}. In isolation, this omission appears minor. In composition, it cascades—when \texttt{digitSum}'s output feeds a downstream function expecting non-negative input, verification fails globally despite both functions being locally correct.
Note that this pattern recurs throughout our dataset. Models generate specifications sufficient for local correctness but insufficient for compositional soundness. A \texttt{requires n >= 0} precondition absent from one function invalidates the entire pipeline's verification, even when each component individually passes most test cases. The fragility stems from a fundamental mismatch: LLMs learn specifications as local patterns rather than global contracts. They lack the architectural machinery to reason about how data constraints propagate through function calls—a capability essential for modular verification.
The implications extend beyond Dafny. Any system requiring compositional correctness—from distributed systems protocols to smart contract verification—will face similar failures until models can reason about specification flow across component boundaries. 

The first key takeaway insight about \textit{specification fragility} is summarized below:  

\begin{tcolorbox}[colback=blue!5!white,colframe=blue!75!black]
  \textbf{Takeaways (\underline{i})}:
  \textit{LLMs handle local specs but fail under composition. Missing contract propagation is the main cause of verification breakdowns.}
\end{tcolorbox}

\subsection{Implementation--Proof Misalignment: The Independence Assumption}
The second failure mode reveals a deeper architectural issue: LLMs treat implementation and specification as independent generation tasks rather than coupled constraints. In 21.7\% of failures, syntactically valid code contradicts its own specifications. One striking example: a model generated \texttt{assert 0 >= 1;} within otherwise reasonable code—not a typo but a systematic failure to maintain logical consistency.
More subtle misalignments prove equally fatal. Loop invariants like \texttt{forall k :: 0 <= k < i ==> cnt[k] >= 0} appear plausible but fail verification because the implementation's array access patterns violate the stated bounds. The model generates invariants that ``look right'' based on training patterns but don't correspond to the actual code behavior. This isn't surprising given transformer architecture: attention mechanisms excel at capturing local dependencies but struggle with the bidirectional constraints between specifications and implementations.
Current training paradigms exacerbate this issue. Models learn from code-specification pairs without explicit feedback on the mutual consistency between them. The result: impressive performance on syntax and moderate success on individual functions, but catastrophic failure when consistency is required across boundaries. 

We summarize the second key takeaway about \textit{implementation-proof misalignment} as:

\begin{tcolorbox}[colback=blue!5!white,colframe=blue!75!black]
  \textbf{Takeaways (\underline{ii})}:
  \textit{Code and specs are generated independently, leading to plausible but inconsistent invariants. Future training for this task should enforce better alignment.}
\end{tcolorbox}

\subsection{Reasoning Instability: Induction as Achilles' Heel}
The third failure pattern, which we refer to as \textit{reasoning instability}, exposes perhaps the most fundamental limitation. Formal verification relies on inductive reasoning: proving properties hold initially, maintain their validity through iterations, and compose across calls. LLMs consistently fail this inductive chain. Loop invariants that should accumulate state (e.g., \texttt{invariant res == stringToIntHelper(s[..i])}) break because models cannot track how program state evolves through iterations. Recursive functions lack proper termination arguments. Properties proven for base cases fail to extend inductively. This instability reflects the inherently statistical nature of reasoning exhibited by transformer architectures. While capable of pattern-matching similar invariants from training data, models cannot construct the inductive proofs verification demands. They approximate rather than prove, which is sufficient for typical NLP tasks but inadequate for verifiable code generation, where formal verification is required. 

We summarize the third insight about \textit{reasoning instability} as:

\begin{tcolorbox}[colback=blue!5!white,colframe=blue!75!black]
  \textbf{Takeaways (\underline{iii})}:
  \textit{LLMs approximate base cases but fail to sustain inductive reasoning, exposing a structural gap in formal verification.}
\end{tcolorbox}

\section{Limitations and Future Work}

While \name exposes fundamental limitations in compositional reasoning, we want to gently mention several constraints of our evaluation, which indicate some interesting future work.

\begin{itemize}[leftmargin=*]
    \item \textbf{Compositional Patterns.} We restrict to chain-based compositions (sequential function calls) rather than complex topologies (recursive compositions, mutual dependencies) due to synthesis tractability. While chains suffice to demonstrate compositional failure, real systems exhibit richer patterns. Extending to arbitrary call graphs requires solving verification tractability for cyclic dependencies—a challenge independent of LLM capabilities.

    \item \textbf{Specification Types.} Our benchmark tests functional correctness (preconditions, postconditions, invariants) but not liveness properties, resource bounds, or security policies. These orthogonal concerns—e.g., proving memory consumption remains constant across compositions—require different verification techniques and evaluation metrics.


    \item \textbf{Data Scarcity.} The core challenge may be training data availability. Repositories contain only a few verified multi-function programs with compositional specifications. Synthetic data generation or bootstrapped program synthesis could address this gap, although ensuring semantic diversity remains a challenge.
\end{itemize}

\section{Conclusion}

We introduce \name, the first benchmark specifically designed to evaluate the generation of compositional specifications for formal verification. Through 300 synthesized multi-function Dafny programs, we systematically assessed 13 state-of-the-art LLMs on their ability to generate specifications that ensure correctness across function boundaries.
Our results reveal a fundamental capability gap: while models achieve greater than 99\% syntax correctness and more than 58\% verification on single-function benchmarks, they collapse to 3.69\% verification on compositional tasks—a 92\% degradation. This performance cliff persists across all model families despite increased sampling (Pass@8), indicating an architectural rather than search limitation. Error analysis identifies three systematic failure modes: specification fragility (39.2\%), implementation-proof misalignment (21.7\%), and reasoning instability (14.1\%), each reflecting the inability to maintain logical commitments across functional boundaries.
In conclusion, \name provides both a diagnostic tool for current systems and a concrete target for future research. We release the benchmark, evaluation framework, and synthesis pipeline to accelerate progress on this critical challenge.

\bibliography{iclr2026_conference}
\bibliographystyle{iclr2026_conference}

\clearpage

\appendix
\section{The Use of LLMs in Writing}
During the preparation of this manuscript, we employed a large language model (OpenAI GPT-5) to assist with language refinement and editorial improvements. Specifically, the LLM was used to enhance sentence fluency, improve clarity of expression, and ensure consistency with academic writing conventions. The tool was applied exclusively for linguistic polishing—all research design, experimental work, data analysis, and core intellectual contributions remain entirely original. 

\section{Broader Impact and Societal Implications}

Our benchmark exposes fundamental gaps in LLM compositional reasoning, underscoring both technical limitations and important societal implications.

\textbf{Positive impacts:} Advancing automated verification could enhance software reliability in safety-critical systems (medical devices, autonomous vehicles) and democratize formal methods for resource-constrained teams. 

\textbf{Risks:} The 92\% verification failure rate we document warns against premature deployment in safety-critical contexts. The specification fragility (39.2\% of failures) is particularly concerning—locally correct but compositionally invalid specifications could create false confidence in system safety. 

\textbf{Accessibility:} While our benchmark is open-source, evaluation requires costly API access to frontier models, potentially creating verification disparities. 

\textbf{Environmental:} Extensive model evaluation across 13 LLMs carries computational costs we acknowledge.

By quantifying the compositional reasoning gap, this work aims to guide development of more reliable verification systems while emphasizing the continued necessity of human expertise in safety-critical software development. This gap may naturally hinder automated malicious system development, while stronger verification could support both good and harmful uses.

\section{Introduction to Dafny}
\label{appendix:intro}
Dafny~\citep{leino2010dafny}, developed at Microsoft Research, is a verification-oriented programming language specifically designed to support formal reasoning about software. Unlike conventional languages where correctness is primarily assessed through testing, Dafny integrates an automated program verifier directly into the development workflow, enabling developers to construct code that is mathematically proven to satisfy its specifications. This approach shifts the discovery of defects from the testing phase to the design and implementation phases, thereby improving software reliability.  

A distinctive feature of Dafny is that specifications are treated as first-class citizens. Methods can be annotated with \emph{preconditions}, \emph{postconditions}, and logical properties that describe intended behavior. For example:  

\begin{lstlisting}[language=Java]
method Example(a: array<int>) returns (b: array<int>)
  requires a.Length > 0                // Precondition
  ensures b.Length == a.Length         // Postcondition  
  ensures forall i :: 0 <= i < b.Length ==> b[i] >= 0  // Property
{
  // Implementation
}
\end{lstlisting}

The Dafny verifier relies on automated theorem proving (via Z3 solver~\citep{de2008z3}) to ensure that implementations conform to these specifications, providing mathematical certainty about program behavior. Crucially, the ability to reason about the composition of verified components determines whether verification can scale from toy examples to real-world systems. Without compositional reasoning, verification remains confined to small, isolated programs rather than production-level software.

\section{Automated Theorem Proving}
\label{appendix:comparison}
A complementary line of work contrasts automated verification frameworks with interactive theorem proving (ITP) systems. Languages such as Dafny and Verus rely on SMT solvers to discharge proof obligations, requiring only lightweight annotations (e.g., invariants, assertions). This design lowers the barrier to entry but is constrained by the solver’s limited reasoning scope and opaque failure modes. In contrast, ITPs such as Lean expose every proof step explicitly, enabling iterative refinement and error diagnosis. Recent studies even show that LLMs can generate competition-level mathematical proofs in Lean. However, existing Lean-based benchmarks (e.g., miniCodeProps, FVAPPS) either focus narrowly on proof synthesis or lack human validation. By comparison, Dafny offers a more balanced environment for benchmarking LLMs: it combines code, specifications, and automated verification in a way that remains close to mainstream programming practice.

\textbf{How Dafny Works and Its Core Strengths.} Dafny's approach stems from its verification-aware design. Developers embed formal specifications, such as preconditions, postconditions, and loop invariants, directly within the code~\citep{leino2010dafny}. These specifications are not merely comments; they are integral components checked by the built-in verifier. The verifier translates Dafny code and its specifications into an intermediate verification language, Boogie, which then generates proof obligations. These obligations are processed by an SMT solver (e.g., Z3) to prove their validity. If all obligations are proven, the code is confirmed to be correct according to its specifications. If a proof fails, Dafny provides precise feedback on the inconsistencies. This methodology supports correctness by construction, helping to reduce common errors like null pointer dereferences or array out-of-bounds access~\citep{poesia2024dafny}. Once verified, Dafny code can be translated into mainstream languages such as Python for execution~\citep{li2025dafny}.

\textbf{Dafny vs. Python: A Fundamental Difference in Approach.} To understand Dafny's position, it's useful to compare it with a widely used language like Python. While both are effective, their fundamental design philosophies and primary objectives differ, as shown in Table~\ref{difference}.

\begin{table}[ht]
\centering
\resizebox{\textwidth}{!}{%
\begin{tabular}{lll}
\toprule
\textbf{Feature}       & \textbf{Dafny}                              & \textbf{Python}                         \\
\midrule
Year Introduced        & 2010 (Microsoft Research)                   & 1991 (Guido van Rossum)                \\
Type System            & Static typing, compile-time checks          & Dynamic typing, run-time checks        \\
Formal Verification    & {\color{ForestGreen}{\textbf{Yes}}} — built-in contracts and proofs         & {\color{red}{No}} — only basic \texttt{assert}        \\
Main Use               & Verified algorithms, critical systems       & General-purpose programming            \\
Execution Model        & Compiled with verification                  & Interpreted (e.g., CPython)            \\
\toprule
\end{tabular}%
}
\caption{Key differences between Dafny and Python.}
\label{difference}
\end{table}

In summary, Dafny offers a distinct approach to software development by integrating formal verification into the language itself. While Python excels in agile development and broad applicability, Dafny is particularly suited for domains where software correctness and formal guarantees are critical. For more, please refer to the Dafny official website\footnote{\url{https://dafny.org/dafny/OnlineTutorial/guide}}.

\section{Prompt for Synthesis}
\label{appendix:prompt_syn}
The prompt templates used for annotating data with Claude 3.5 Sonnet are shown in the following boxes.

\begin{tcolorbox}[
    width=\textwidth,
    colframe=black,
    colback=white,
    boxrule=1pt,
    arc=2mm,
    title={\textbf{Prompt for Inital Dafny Code Generation}},
    fonttitle=\bfseries\large,
    coltitle=white,
    colbacktitle=black,
    center title,
    bottom=1mm,
    top=1mm,
    boxsep=3mm,
    before skip=10pt,
    after skip=15pt,
]

\textbf{\color{slategray}SYSTEM}

You are an expert AI assistant that writes Dafny programs. You excel at writing code with formally verified correctness, providing precise preconditions and postconditions, and finding the appropriate loop invariants to ensure all verification conditions are met.

\vspace{6pt}
\textbf{\color{slategray}TASK}

Below is the Python code:
\begin{tcolorbox}[
    colback=gray!10,
    colframe=gray!50,
    boxrule=0.2pt,
    left=2pt,
    right=2pt,
    top=1pt,
    bottom=1pt,
    sharp corners,
    width=\textwidth,
]
\texttt{\textasciigrave\textasciigrave\textasciigrave python} \\
\texttt{<python\_code>} \\
\texttt{\textasciigrave\textasciigrave\textasciigrave}
\end{tcolorbox}

Please translate this Python code into Dafny, ensuring:

\begin{enumerate}
    \item \textbf{Method Signatures}: Each piece of functionality should be expressed as a Dafny method (or set of methods) with a well-defined signature.
    \item \textbf{Preconditions}: Clearly state any `requires` clauses for each method (e.g., array length constraints, non-null references, numeric domain restrictions, etc.).
    \item \textbf{Postconditions}: State the logical guarantees about the returned values or final state as `ensures` clauses (e.g., correctness of returned results, absence of side effects, etc.).
    \item \textbf{Verification Details}: Include all necessary loop invariants (or other verification hints) so Dafny can prove the postconditions, along with a brief explanation. For example:
    - Explain how you chose your invariants.
    - Describe how they ensure the correctness of the loop.
\end{enumerate}

Return the final Dafny code as a self-contained snippet that can be verified by Dafny as-is, with a short explanation of how it connects to the original Python functionality.

\vspace{6pt}
\textbf{\color{slategray}AI ASSISTANT}

$<$The LLM's generated Dafny code with specifications here.$>$
\label{fig:prompt_2}
\end{tcolorbox}

\begin{tcolorbox}[
    width=\textwidth,
    colframe=black,
    colback=white,
    boxrule=1pt,
    arc=2mm,
    title={\textbf{Dynamic Debugging Prompt for Code Generation}},
    fonttitle=\bfseries\large,
    coltitle=white,
    colbacktitle=black,
    center title,
    bottom=1mm,
    top=1mm,
    boxsep=3mm,
    before skip=10pt,
    after skip=15pt,
]

\textbf{\color{slategray}SYSTEM}

You are an expert AI assistant that writes and debugs Dafny programs. You excel at diagnosing and fixing verification errors based on Dafny solver messages, while maintaining correct preconditions, postconditions, and loop invariants.

\vspace{6pt}
\textbf{\color{slategray}TASK}

Below is the Python code:
\begin{tcolorbox}[
    colback=gray!10,
    colframe=gray!50,
    boxrule=0.2pt,
    left=2pt,
    right=2pt,
    top=1pt,
    bottom=1pt,
    sharp corners,
    width=\textwidth,
]
\texttt{\textasciigrave\textasciigrave\textasciigrave python} \\
\texttt{<python\_code>} \\
\texttt{\textasciigrave\textasciigrave\textasciigrave}
\end{tcolorbox}

And the Dafny code you previously provided (which I tried to verify):
\begin{tcolorbox}[
    colback=gray!10,
    colframe=gray!50,
    boxrule=0.2pt,
    left=2pt,
    right=2pt,
    top=1pt,
    bottom=1pt,
    sharp corners,
    width=\textwidth,
]
\texttt{\textasciigrave\textasciigrave\textasciigrave dafny} \\
\texttt{<main\_spec>} \\
\texttt{\textasciigrave\textasciigrave\textasciigrave}
\end{tcolorbox}

I ran \texttt{dafny verify *.dfy} and received this error message:
\begin{tcolorbox}[
    colback=gray!10,
    colframe=gray!50,
    boxrule=0.2pt,
    left=2pt,
    right=2pt,
    top=1pt,
    bottom=1pt,
    sharp corners,
    width=\textwidth,
]
\texttt{\textasciigrave\textasciigrave\textasciigrave} \\
\texttt{<dafny\_analysis\_result>} \\
\texttt{\textasciigrave\textasciigrave\textasciigrave}
\end{tcolorbox}

Can you please fix the main function specification so that it parses successfully? Output the corrected main function specification only, without any other text.

\vspace{6pt}
\textbf{\color{slategray}AI ASSISTANT}

$<$The LLM's generated Dafny code with specifications here.$>$
\label{fig:prompt_1}
\end{tcolorbox}

\section{Data Contamination Analysis}
\label{sec:contamination}
To validate the novelty of \name, we conducted a rigorous data contamination analysis against the widely-used \textsc{MBPP} dataset~\citep{austin2021program}, used to assess contamination in Python source data. We confirm that our benchmark source data shows no significant overlap, ensuring model performance reflects genuine reasoning capabilities rather than memorization.

Our analysis, focusing solely on code, employs two standard metrics: \textbf{Exact Match} to detect verbatim copies, and \textbf{n-gram Jaccard Similarity} to identify structurally similar code. We performed this analysis under four distinct configurations, the results of which are summarized in Table~\ref{tab:contamination_final}.

Across all scenarios, we found \textbf{zero exact matches}. The n-gram Jaccard similarity remains negligible, peaking at a mere 0.0078 even under the most aggressive settings. These findings provide strong evidence that \name is free from training data contamination.

\begin{table}[ht]
\centering
\caption{
    Summary of Data Contamination Analysis. The table shows results for four testing configurations: 
    \textbf{A (Conservative)} with minimal preprocessing; 
    \textbf{B (Default)} with moderate preprocessing; 
    \textbf{C (Aggressive)} with extensive preprocessing; and 
    \textbf{D (Holistic)} for a structure-level check. 
    Across all configurations, results show zero exact matches and negligible n-gram similarity when comparing \name source data against \textsc{MBPP}, confirming the benchmark's integrity.
}
\label{tab:contamination_final}
\begin{tabular}{lcccc}
\toprule
\textbf{Analysis Configuration} & \textbf{N-gram (\(n\))} & \textbf{Exact Overlap} & \textbf{Max Jaccard} & \textbf{vs. sanitized-mbpp} \\
\midrule
A: Conservative & 15 & 0 & 0.000078 & 0 \\
B: Default      & 11, 13, 15 & 0 & 0.000389 & 0 \\
C: Aggressive     & 9, 11, 13  & 0 & 0.007757 & 0 \\
D: Holistic       & 11, 13, 15 & 0 & 0.000234 & 0 \\
\bottomrule
\end{tabular}
\end{table}

\section{Prompt for Evaluation}
\label{appendix:prompt}
The prompt template used for evaluation is shown in the following box. Note that all model outputs are used directly for Dafny verification.
\begin{tcolorbox}[
    width=\textwidth,
    colframe=black,
    colback=white,
    boxrule=1pt,
    arc=2mm,
    title={\textbf{Evaluation Prompt for Dafny Specification Generation}},
    fonttitle=\bfseries\large,
    coltitle=white,
    colbacktitle=black,
    center title,
    bottom=1mm,
    top=1mm,
    boxsep=3mm,
    before skip=10pt,
    after skip=15pt,
]

\textbf{\color{slategray}SYSTEM}

You are an expert in Dafny. You will be given tasks dealing with Dafny programs including precise annotations. You should only return code body in all circumstances. No text is allowed.

\vspace{6pt}
\textbf{\color{slategray}TASK}

Given a Dafny program with function signature, preconditions, postconditions, and code, but with annotations missing. Please return a complete Dafny program with the strongest possible annotation (loop invariants, assert statements, etc.) filled back in. Do not explain or output any text. If you have to explain, put all explanations in comments form. There should only be code body in your output. Please use exactly the same function signature, preconditions, and postconditions. Do not ever modify the given lines.

Below is the program:
\begin{tcolorbox}[
    colback=gray!10,
    colframe=gray!50,
    boxrule=0.2pt,
    left=2pt,
    right=2pt,
    top=1pt,
    bottom=1pt,
    sharp corners,
    width=\textwidth,
]
\texttt{\textasciigrave\textasciigrave\textasciigrave dafny} \\
\texttt{<dafny\_program\_with\_missing\_annotations>} \\
\texttt{\textasciigrave\textasciigrave\textasciigrave}
\end{tcolorbox}

\vspace{6pt}
\textbf{\color{slategray}AI ASSISTANT}

\begin{tcolorbox}[
    colback=gray!10,
    colframe=gray!50,
    boxrule=0.2pt,
    left=2pt,
    right=2pt,
    top=1pt,
    bottom=1pt,
    sharp corners,
    width=\textwidth,
]
\texttt{\textasciigrave\textasciigrave\textasciigrave dafny} \\
\texttt{<The LLM's generated Dafny code with specifications here.>} \\
\texttt{\textasciigrave\textasciigrave\textasciigrave}
\end{tcolorbox}

\end{tcolorbox}

\clearpage

\section{Real Examples from Benchmark Data}
\label{appendix:real_examples}

\subsection{Challenge 1: Formal Specification Synthesis}

The following examples demonstrate failures in generating correct formal specifications, particularly preconditions and postconditions.

\subsubsection{Example 1.1: Postcondition Failure (org\_input\_id: 2, self\_id: 20001)}
\label{appendix:example11}

\textbf{LLM Response:}
\begin{lstlisting}
method digitSum(n: int) returns (sum: int)
    requires n >= 0
    ensures sum >= 0
{
    var temp := n;
    sum := 0;
    while temp > 0
        invariant temp >= 0
        invariant sum >= 0
    {
        sum := sum + (temp % 10);
        temp := temp / 10;
    }
}
\end{lstlisting}

\textbf{Dafny Error:}
\begin{lstlisting}
ex.dfy(137,0): Error: a postcondition could not be proved on this return path
ex.dfy(136,25): Related location: this is the postcondition that could not be proved
\end{lstlisting}

\subsubsection{Example 1.2: Missing Precondition (org\_input\_id: 6, self\_id: 60005)}
\label{appendix:example12}
\textbf{LLM Response:}
\begin{lstlisting}
method DigitSum(n: int) returns (sum: int)
    requires n >= 0
    ensures sum >= 0
{
    var num := n;
    sum := 0;
    while num > 0
        invariant sum >= 0
        invariant num >= 0
        decreases num
    {
        sum := sum + (num % 10);
        num := num / 10;
    }
}
\end{lstlisting}

\textbf{Dafny Error:}
\begin{lstlisting}
ex.dfy(177,36): Error: a precondition for this call could not be proved
\end{lstlisting}

\subsection{Challenge 2: Ensuring Code Correctness}

These examples show implementation issues where the code logic doesn't satisfy the specifications.

\subsubsection{Example 2.1: Array Bounds Error (org\_input\_id: 1, self\_id: 10002)}
\label{appendix:example21}
\textbf{LLM Response:}
\begin{lstlisting}
method smallestNumber_2165(num: int) returns (result: int)
{
    var neg := num < 0;
    var n := if num < 0 then -num else num;
    var cnt := new int[10];
    var i := 0;
    while i < 10
        invariant 0 <= i <= 10
        invariant cnt.Length == 10
        invariant forall k :: 0 <= k < i ==> cnt[k] >= 0
    {
        cnt[i] := 0;
        i := i + 1;
    }
    // ... rest of implementation
}
\end{lstlisting}

\textbf{Dafny Error:}
\begin{lstlisting}
ex.dfy(44,30): Error: this loop invariant could not be proved on entry
Related message: loop invariant violation
\end{lstlisting}

\subsubsection{Example 2.2: Assertion Failure (org\_input\_id: 6, self\_id: 60001)}
\label{appendix:example22}
\textbf{LLM Response:}
\begin{lstlisting}
method DigitSum(n: int) returns (sum: int)
    requires n >= 0
    ensures sum >= 0
{
    var num := n;
    sum := 0;
    while num > 0
        invariant sum >= 0
        invariant num >= 0
        decreases num
    {
        sum := sum + (num % 10);
        num := num / 10;
    }
}
\end{lstlisting}

\textbf{Dafny Error:}
\begin{lstlisting}
ex.dfy(181,4): Error: assertion might not hold
assert o2 >= 1;
\end{lstlisting}

\clearpage

\subsection{Challenge 3: Proof Generation and Formal Reasoning}

Examples demonstrating failures in generating proper loop invariants and formal proofs.

\subsubsection{Example 3.1: Invariant Not Maintained (org\_input\_id: 0, self\_id: 6)}
\label{appendix:example31}
\textbf{LLM Response:}
\begin{lstlisting}
method stringToInt(s: seq<char>) returns (result: int)
requires forall i :: 0 <= i < |s| ==> '0' <= s[i] <= '9'
ensures result >= 0
{
    var res := 0;
    var i := 0;
    while i < |s|
        invariant 0 <= i <= |s|
        invariant res >= 0
        invariant res == stringToIntHelper(s[..i])
    {
        res := res * 10 + digitToInt(s[i]);
        i := i + 1;
    }
    result := res;
}
\end{lstlisting}

\textbf{Dafny Error:}
\begin{lstlisting}
ex.dfy(25,18): Error: this invariant could not be proved to be maintained by the loop
Related message: loop invariant violation
\end{lstlisting}

\subsubsection{Example 3.2: Recursion Termination (org\_input\_id: 4, self\_id: 40003)}
\label{appendix:example32}
\textbf{LLM Response:}
\begin{lstlisting}
function gcd_func(a: int, b: int): int
{
    if b == 0 then a else gcd_func(b, a % b)
}

method gcd(a: int, b: int) returns (result: int)
requires a > 0 && b >= 0
ensures result == gcd_func(a, b)
{
    var x := a;
    var y := b;
    while y != 0
        invariant x > 0 && y >= 0
        invariant gcd_func(a, b) == gcd_func(x, y)
        decreases y
    {
        var temp := x % y;
        x := y;
        y := temp;
    }
    result := x;
}
\end{lstlisting}

\textbf{Dafny Error:}
\begin{lstlisting}
ex.dfy(3,34): Error: cannot prove termination; try supplying a decreases clause
if b == 0 then a else gcd_func(b, a % b)
\end{lstlisting}

\subsection{Challenge 4: Spec-Code-Proof Alignment}

These examples show misalignment between specifications and implementation, particularly method vs. function usage.

\subsubsection{Example 4.1: Method in Expression Context (org\_input\_id: 0, self\_id: 5)}

\textbf{LLM Response:}
\begin{lstlisting}
method stringToInt(s: seq<char>) returns (result: int)
requires forall i :: 0 <= i < |s| ==> '0' <= s[i] <= '9'
ensures result >= 0
{
    var res := 0;
    var i := 0;
    while i < |s|
        invariant 0 <= i <= |s|
        invariant res >= 0
        invariant res == if i == 0 then 0 else stringToInt(s[..i])
    {
        res := res * 10 + digitToInt(s[i]);
        i := i + 1;
    }
    result := res;
}
\end{lstlisting}

\textbf{Dafny Error:}
\begin{lstlisting}
ex.dfy(26,43): Error: expression is not allowed to invoke a method (stringToInt)
invariant res == if i == 0 then 0 else stringToInt(s[..i])
\end{lstlisting}

\clearpage

\subsubsection{Example 4.2: Function vs Method Confusion (org\_input\_id: 16, self\_id: 160000)}

\textbf{LLM Response:}
\begin{lstlisting}
method reverse_7(x: int) returns (ans: int)
{
    ans := 0;
    var temp_x := x;
    while temp_x != 0
        invariant ans >= 0
        decreases if temp_x >= 0 then temp_x else -temp_x
    {
        ans := ans * 10 + temp_x % 10;
        temp_x := temp_x / 10;
    }
}

method isBeautiful(n: int) returns (result: bool)
{
    // Implementation for checking if number is beautiful
    result := true; // Simplified
}

method nextBeautifulNumber_769(n: int) returns (result: int)
{
    var x := n + 1;
    while true
        invariant x >= n + 1
        invariant forall y :: n + 1 <= y < x ==> !isBeautiful(y)
        decreases *
    {
        var beautiful := isBeautiful(x);
        if beautiful {
            result := x;
            return;
        }
        x := x + 1;
    }
}
\end{lstlisting}

\textbf{Dafny Error:}
\begin{lstlisting}
ex.dfy(248,46): Error: expression is not allowed to invoke a method (isBeautiful)
invariant forall y :: n + 1 <= y < x ==> !isBeautiful(y)
\end{lstlisting}

\clearpage
\subsection{Challenge 5: Complexity and Scalability Issues}

Examples where complex programs or large data structures cause verification failures.

\subsubsection{Example 5.1: Large Array Allocation (org\_input\_id: 1, self\_id: 10002)}

\textbf{LLM Response:}
\begin{lstlisting}
method smallestNumber_2165(num: int) returns (result: int)
{
    var neg := num < 0;
    var n := if num < 0 then -num else num;
    var cnt := new int[10];
    var i := 0;
    while i < 10
        invariant 0 <= i <= 10
        invariant cnt.Length == 10
        invariant forall k :: 0 <= k < i ==> cnt[k] >= 0
    {
        cnt[i] := 0;
        i := i + 1;
    }

    // Count digits
    var temp := n;
    while temp > 0
        invariant temp >= 0
        invariant forall k :: 0 <= k < 10 ==> cnt[k] >= 0
        decreases temp
    {
        var digit := temp % 10;
        cnt[digit] := cnt[digit] + 1;
        temp := temp / 10;
    }

    // Construct result
    result := 0;
    i := if neg then 1 else 0;
    while i < 10
        invariant 0 <= i <= 10
        invariant result >= 0
        decreases 10 - i
    {
        while cnt[i] > 0
            invariant cnt[i] >= 0
            invariant result >= 0
            decreases cnt[i]
        {
            result := result * 10 + i;
            cnt[i] := cnt[i] - 1;
        }
        i := i + 1;
    }

    if neg {
        result := -result;
    }
}
\end{lstlisting}

\textbf{Dafny Error:}
\begin{lstlisting}
ex.dfy(44,30): Error: this loop invariant could not be proved on entry
Related message: loop invariant violation
invariant ans >= 0
\end{lstlisting}

\subsection{Challenge 7: Avoiding Trivial Solutions}

\textbf{Note:} This category had limited representation in the dataset. The theoretical examples that would fit this category include:

\begin{itemize}
\item Overly permissive postconditions like \texttt{ensures true}
\item Methods with missing functional specifications
\item Weak preconditions that don't prevent edge cases
\item Specifications that allow degenerate solutions
\item Missing bounds on return values
\item Incomplete coverage of input space
\item Trivial loop invariants that don't capture semantics
\item Weak assertions that don't prevent bugs
\item Generic error handling without specific constraints
\end{itemize}

\subsection{Challenge 8: Limited Training in Formal Domains}

Examples showing basic syntax errors and misuse of Dafny constructs.

\subsubsection{Example 8.1: Parse Error (org\_input\_id: 0, self\_id: 0)}

\textbf{LLM Response:}
\begin{lstlisting}
function digitToInt(c: char): int
  requires '0' <= c <= '9'
  ensures 0 <= digitToInt(c) <= 9
{
  (c as int) - ('0' as int)
}

function intToChar(i: int): char
  requires 0 <= i <= 9
  ensures '0' <= intToChar(i) <= '9'
{
  (('0' as int) + i) as char
}

method stringToInt(s: string) returns (result: int)
  requires forall i :: 0 <= i < |s| ==> '0' <= s[i] <= '9'
  ensures result >= 0
{
  var res := 0;
  var i := 0;
  while i < |s|
    invariant 0 <= i <= |s|
    invariant res >= 0
  {
    res := res * 10 + digitToInt(s[i]);
    i := i + 1;
  }
  result := res;
}
\end{lstlisting}

\textbf{Dafny Error:}
\begin{lstlisting}
ex.dfy(16,35): Error: type string is not expected in Dafny; use seq<char> instead
method stringToInt(s: string) returns (result: int)

2 parse errors detected in ex.dfy
\end{lstlisting}

\subsubsection{Example 8.2: Syntax Error (org\_input\_id: 0, self\_id: 2)}

\textbf{LLM Response:}
\begin{lstlisting}
function digitToInt(c: char): int
  requires '0' <= c <= '9'
  ensures 0 <= digitToInt(c) <= 9
{
  (c as int) - ('0' as int)
}

function intToChar(i: int): char
  requires 0 <= i <= 9
  ensures '0' <= intToChar(i) <= '9'
{
  (('0' as int) + i) as char
}

method stringToInt(s: seq<char>) returns (result: int)
  requires forall i :: 0 <= i < |s| ==> '0' <= s[i] <= '9'
  ensures result >= 0
{
  var res := 0;
  var i := 0;
  while i < |s|
    invariant 0 <= i <= |s|
    invariant res >= 0
  {
    res := res * 10 + digitToInt(s[i]);
    i := i + 1;
  }
  result := res;
}
\end{lstlisting}

\textbf{Dafny Error:}
\begin{lstlisting}
ex.dfy(29,5): Error: "closeparen" expected
result := res;

2 parse errors detected in ex.dfy
\end{lstlisting}

\end{document}